\documentclass[aps,prl,superscriptaddress,showpacs]{revtex4}
\usepackage{graphicx}
\usepackage{bm}

\newcommand{\beq}{\begin{equation}}
\newcommand{\eeq}{\end{equation}}
\newcommand{\bea}{\begin{eqnarray}}
\newcommand{\eea}{\end{eqnarray}}

\newcommand{\aks}{\hat{a}_{{\bf k},s}}

\newcommand{\emk}{\,e^{i \left( {\bf k} \cdot {\bf r} - c|{\bf k}| t  \right)}}

\newcommand{\bfk}{{\bf k}}
\newcommand{\eks}{{\bf e}_{{\bf k},s}}
\newcommand{\kperp}{{\bf k}_{\perp}}
\newcommand{\Dto}{\Delta_{\theta_0} (\bfk)}

\begin{document}

\title{Quantum electromagnetic X-waves}

\author{Alessandro Ciattoni} \email{alessandro.ciattoni@aquila.infn.it}
\affiliation{Consiglio Nazionale delle Ricerche, CASTI Regional Lab 67010 L'Aquila, Italy} \affiliation{Dipartimento di Fisica,
Universit\'a dell'Aquila, 67010 L'Aquila, Italy}

\author{Claudio Conti} \email{claudio.conti@phys.uniroma1.it}
\affiliation{Research Center "Enrico Fermi" and SOFT INFM-CNR, Universit\'a La Sapienza, Piazzale Aldo Moro 2, 00185 Rome, Italy}

\date{\today}
\begin{abstract}
We show that two distinct quantum states of the electromagnetic field can be associated to a classical vector X wave or a
propagation-invariant solution of Maxwell equations. The difference between the two states is of pure quantum mechanical origin
since they are internally entangled and disentangled, respectively and can be generated by different linear or nonlinear
processes. Detection and generation of Schr\"odinger-cat states comprising two entangled X-waves and their possible applications
are discussed.
\end{abstract}


\maketitle

\section{Introduction}

Electromagnetic X waves and, more generally, ``localized waves'' \cite{RecamiBook} are propagation-invariant solutions of
classical Maxwell equations in vacuum \cite{Lu92,Ziolkowski89,ciattoni04b}, dielectric media \cite{Sonajalg96}, plasmas
\cite{Ciattoni04d}, optically nonlinear materials \cite{Berge02,Orlov02,Conti03,Staliunas05} and periodic structures
\cite{Conti04,Longhi04,Manela05,Lahini07}. In recent years these waves have attracted a considerable interest since, despite at a
first glance they are non-physical (for example some of them apparently travel superluminally and carry an infinite amount of
energy), propagation-invariant fields support several applications like image transmission, reconstruction and telecommunications
(for a recent review see Ref.\cite{RecamiBook}).

Since the first observation of localized optical X waves \cite{Saari97}, the fundamental implications of rigidly travelling
spatio-temporal correlations have been recognized. Classically, these waves can be generated by means of linear devices (like
axicons) \cite{Saari97} and, recently, their experimental feasibility has been also proved through nonlinear optical processes
\cite{Jedrkiewicz06,Picozzi02}. Within the quantum framework, analytical techniques borrowed from the subject of localized waves
have allowed some authors to consider highly localized states for the photon-wavefunction \cite{Saari05}, whereas the role of
quantum X waves in nonlinear optical process (limited to the paraxial regime) were analyzed in \cite{Conti04d}.

In this paper we consider the issue of ``constructing a quantum state of the electromagnetic field whose classical counterpart is
an X-wave satisfying the full set of Maxwell equations in vacuum''. More precisely, we investigate the possibility of introducing
various quantum states on which the mean value of the electric field operator coincides with a prescribed classical vector X
wave. We show that the solution of this problem is not-trivial by finding two different quantum states characterized by the above
property. Even if from a classical perspective they describe the same entity, these two states essentially differ in the degree
of entanglement of their internal modal structure so that both a ``disentangled'' and  an ``entangled'' quantum description can
be provided of an arbitrary classical X wave. In this respect one can investigate a method for discerning between the two
possibilities and, correspondingly, for recognizing which kind of quantum X wave is produced by a given mechanism (e.g. axicons
or nonlinear processes). In addition, we show that some X wave states can be considered as multi-mode, multi-dimensional
Schr\"odinger cats, exhibiting the relevant feature of being propagation invariant; the corresponding macroscopic quantum content
is controlled by the velocity of the X waves and can be maintained for long distances.

\section{Quantum Description of X waves}

Within a classical formulation, the $x$-component of the electric field of a radially symmetric linearly polarized X wave can be
written as (extension to more general X waves solutions \cite{RecamiBook} is trivial)
\begin{equation} \label{CX}
E_x ({\bf r},t) =\Re \int_0^\infty dk f(k) J_0(\sin \theta_0 k r_\perp) e^{i k \cos \theta_0 \left(z- \frac{c}{\cos \theta_0} t
\right)}
\end{equation}
where $r_\perp = \sqrt{x^2+y^2}$ and $\Re$ stands for the real part, whereas the longitudinal component is simply expressed in
terms of the transversal components. (see e.g. \cite{ciattoni04b}) This kind of classical waves propagates along the $z-$axis
without distortion and it is fully characterized by its spectrum $f(k)$ and its velocity $c/\cos \theta_0$. Physically, such an X
wave arises as the superposition of various plane-waves at different frequencies whose wave-vectors are all inclined at the same
angle $\theta _0$ with the propagation direction $z$. In order to give a quantum description of X waves, we construct a quantum
state of the electromagnetic field through the requirement that the quantum mean value of the electric field operator on this
state yields a propagation invariant wave of the kind of Eq.(\ref{CX}) and we show that this can be done in two different
manners. In order to fix the notation, we start from the standard expression of the electric field operator in Heisenberg
representation, $\hat{\bf E} ({\bf r},t) = \sum_{{\bf k},s} \sqrt{\frac{\hbar c |{\bf k}|}{2\varepsilon_0 L^3}} i \aks \emk \eks
+ \text{h.c.}$, where $L$ is the edge of the quantization cubic box (see e.g. \cite{MandelBook,LoudonBook}), ${\bf k} = (2 \pi
/L) (n_x {\bf e}_x + n_y {\bf e}_y +n_z {\bf e}_z)$ (where $n_x$, $n_y$ and $n_z$ are integers), $\eks$ are the pair ($s=1,2$) of
polarization unit vectors associated to the mode $\bf k$ and $\aks$ are the standard photon annihilation operators. Bearing in
mind that the mean value of the electric field operator on a coherent state is a plane wave, let us consider the quantum states
\begin{equation} \label{X1}
|X \rangle=\prod_{\bfk,s}\hat{D}_{\bfk,s} [\Delta_{\theta_0}(\bfk) \beta(\bfk,s)]|0\rangle
\end{equation}
where $\hat{D}_{\bfk,s}(\alpha)$ is the displacement operator which, acting on the vacuum state $|0\rangle$, produces the
coherent state $|\alpha \rangle_{\bfk,s}$ (associated to the mode ${\bfk,s}$) according to the relation $\hat{D}_{\bfk,s}(\alpha)
|0\rangle = |\alpha \rangle_{\bfk,s}$ \cite{MandelBook}. Here $\beta(\bfk,s)$ is a complex weight function to be determined below
and, in order to deal with the discreteness of the $\bf k$ vectors, we have defined $\Dto=1$ if $\bf k \in \mathcal{C} (\theta_0
,\delta \theta)$ and $\Dto=0$ elsewhere, where $\mathcal{C}(\theta_0,\delta \theta)$ is the portion of space comprised between
the two coaxial cones of aperture angles $\theta_0-\delta\theta/2$ and $\theta_0+\delta\theta/2$ and whose common axis coincides
with the $z-$axis. The presence of the function $\Dto$ assures that the state $|X\rangle$ of Eq.(\ref{X1}) is the tensor product
of coherent states whose modes have wave vectors globally lying, for $\delta \theta \ll \theta_0$, on a cone with aperture angle
$\theta_0$ so that the relation $k_z=\eta |\kperp|$ with $\eta=1/\tan \theta_0$ holds for each wave vector. The mean value of the
electric field operator on the state $|X\rangle$ is given by
\beq \label{averE} \langle X| \hat{\bf E}| X \rangle = \displaystyle \Re \left[ \sum_{\bfk,s} \Delta_{\theta_0}(\bfk)
\sqrt{\frac{2 \hbar c |\bfk|}{\varepsilon_0 L^3}} i \beta(\bfk,s) \emk \eks \right] \eeq
from which we envisage that, for $L \rightarrow \infty$ and $\delta \theta \ll \theta_0$, $ \langle X |\hat{\bf E} ({\bf r},t ) |
X \rangle \rightarrow {\bf E}(x,y,z-V t)$, which is a genuine propagation-invariant vector field travelling along the $z-$axis.
In order to prove this assertion we note that the limit is different from zero only if $\beta$ is dependent on $\delta \theta$
and $L$ and it is given by
\beq \label{beta} \beta({\bf k},s)=\displaystyle \frac{(2 \pi)^3}{ i \delta \theta}\sqrt{\frac{\varepsilon_0}{2\hbar c |\bfk|
L^3}} \frac{|\bfk|}{|\kperp|}f_s(\bfk) \text{,} \eeq
where ${\bf k}_\perp = k_x {\bf e}_x + k_y {\bf e}_y$, $f_s(\bfk)$ are two arbitrary function of $\bf k$ and the various
coefficients are chosen for later convenience. Substituting Eq.(\ref{beta}) into Eq.(\ref{averE}) and performing the limits $L
\rightarrow +\infty$ and $\delta \theta\rightarrow 0$ by means of the rule $\frac{1}{L^3} \sum_{\bfk} \rightarrow
\frac{1}{(2\pi)^3} \int d^3 \bfk $ and the relation $\lim_{\delta \theta \rightarrow 0} \Dto / \delta \theta = \delta (\theta -
\theta_0)$ we obtain
\begin{equation} \label{XX}
\langle X |  \hat{\bf E} |X \rangle = \displaystyle \Re \left\{ \int_0^\infty dk k^2 \int_0^{2\pi} d\phi
 \left[f_1({\bf k}) {\bf e}_{{\bf k} 1} + f_2({\bf k}) {\bf e}_{{\bf k} 2} \right] e^{i ( {\bf k} \cdot {\bf r} - c k t)} \right\}
\end{equation}
where, after defining ${\bf k}= k[\sin \theta_0 (\cos \phi {\bf e}_x + \sin \phi {\bf e}_y ) + \cos \theta_0 {\bf e}_z ]$, polar
coordinates has been introduced for the $\bf k$ integration. Note that, inside the integral of Eq.(\ref{XX}), the relation ${\bf
k} \cdot {\bf r} - c k t = k [(x \cos \phi \sin \theta_0  + y \sin \phi \sin \theta_0) +  \cos \theta_0 (z - ct / \cos \theta_0
)]$ holds, so that the field in Eq.(\ref{XX}) describes a wave travelling undistorted along the $z-$axis with velocity $V=c/\cos
\theta_0$, as expected. Note that, strictly speaking, this rigorous nondiffracting behavior is a consequence of the limit $\delta
\theta \rightarrow 0$ which is an idealization of the realistic condition $\delta \theta \ll \theta_0$, $\delta \theta$
representing the experimental uncertainty of the wave vectors smeared around the selected conical surface. Since $\delta \theta$
can be experimentally chosen much smaller than $\theta _0$, the actual fields are nearly nondiffracting objects, their departure
from ideal fields being experimentally tunable.

A convenient choice for the modes polarization unit vectors is given by ${\bf e}_{{\bf k} 1} = \cos \theta_0 (\cos \phi {\bf e}_x
+ \sin \phi {\bf e}_y ) - \sin \theta_0 {\bf e}_z $ and ${\bf e}_{{\bf k} 2} = -\sin \phi {\bf e}_x + \cos \phi {\bf e}_y $ so
that a linearly polarized wave along the $x-$direction is obtained by letting $f_2=-f_1 \tan \phi \cos \theta_0$ with $f_1= (
\cos \phi / \cos \theta_0) [f(k)/k^2]$ from which we get $f_1({\bf k}) {\bf e}_{{\bf k} 1} + f_2({\bf k}) {\bf e}_{{\bf k} 2} =
f(k) \left( {\bf e}_x -\cos \phi \tan \theta_0 {\bf e}_z \right) / k^2$ where the factor $k^{-2}$ has been added for later
convenience. Inserting this expression into Eq.(\ref{XX}), we obtain
\begin{equation} \label{LX}
\langle X | \hat{\bf E} | X\rangle = \Re \int_0^\infty dk f(k) \left[J_0 \left( \sin \theta_0 k r_{\perp} \right) {\bf e}_x -
\frac{x \tan \theta_0}{r_\perp} J_1 \left(\sin \theta_0 k r_{\perp} \right) {\bf e}_z \right] e^{i k \cos \theta_0 (z- V t)}
\end{equation}
where $J_n (\xi)$ is the Bessel function of the first kind of order n and the relation $\int_0^{2\pi} d \phi e^{i h (x \cos\phi +
y \sin\phi)} = 2\pi J_0 (h r_{\perp})$ has been exploited. The expectation value of the electric field in Eqs.(\ref{LX})
describes a linearly polarized X-waves (see Eq.(\ref{CX})) which is an exact solution of Maxwell equations in vacuum (see
Ref.\cite{ciattoni04b}). Note that the longitudinal component $E_z$ is due to the full non-paraxial character of the exact
approach we are considering and it is negligible for small aperture angles $\theta_0$.

Let us consider the state of the electromagnetic field given by
\beq \label{XE}
|X_E \rangle = \sum_{{\bf k},s} \Dto \Phi(\bfk,s) \hat{D}_{\bfk,s}\left[\alpha(\bfk,s)\right]|0\rangle \eeq
where $\alpha(\bfk,s)$ and $\Phi(\bfk,s)$ are arbitrary complex function, the second one being constrained by the normalization
conditions $\langle X_E | X_E \rangle=1$. Note that the very presence of the function $\Dto$ in Eq.(\ref{XE}) implies that modes
are collected in such a way that their wave vectors almost lies, in the limit $\delta \theta \ll \theta_0$, on the surface of the
cone characterizing the spectrum of an arbitrary classical X wave and, therefore $|X_E \rangle$ is expected to describe a quantum
X wave. In order to prove this assertion we note that
\begin{eqnarray}
\langle X_E| \hat{a}_{\bfk s} |X_E\rangle = \Dto \alpha(\bfk,s) Q(\bfk,s)
\end{eqnarray}
where $Q(\bfk,s)=|\Phi(\bfk,s)|^2 + \Phi(\bfk,s) e^{-\frac{1}{2} |\alpha(\bfk,s)|^2} \displaystyle \sum_{{\bf k'}\neq {\bf k} ,
s'\neq s} \Delta_{\theta_0}(\bfk') \Phi^*(\bfk',s') e^{-\frac{1}{2} |\alpha(\bfk',s')|^2}$ so that the mean value of the electric
field on $|X_E \rangle$ turns out to be
\beq \label{avexe} \langle X_E | \hat{\bf E} | X_E\rangle= \Re \left[ \displaystyle\sum_{\bfk,s} \Dto G(\bfk,s)
e^{i\left(\bfk\cdot {\bf r}-c |\bfk|t\right)}\eks \right] \eeq
where $G(\bfk,s)= i\sqrt{\frac{2\hbar c |\bfk|}{\varepsilon_0 L^3}} \alpha(\bfk,s) Q(\bfk,s)$. Comparing Eq.(\ref{avexe}) with
Eq.(\ref{averE}) and noting that their structures are identical, we deduce that, in the limit $\delta \theta \rightarrow 0$ and
$L \rightarrow \infty$, the state $|X_E\rangle$ describes a quantum X wave. It is worth noting that this state is completely
different from the $| X \rangle$ of Eq.(\ref{X1}) and it is obtained as a linear combination of the states
$\hat{D}_{\bfk,s}[\alpha(\bfk,s)]|0\rangle$ each corresponding to a coherent state in the mode $(\bfk,s)$, all other modes being
in the vacuum state. This implies that the state $|X_E \rangle$ arises from an entangled superposition of modes in coherent
states, where with ``entanglement'' we mean that the state cannot be expressed as a product of kets containing different modes.
We therefore conclude that a prescribed classical X wave can be represented by two different quantum states $|X\rangle$ and $|X_E
\rangle$ exhibiting an internal disentangled and entangled modal structure, respectively whose quantum difference is expected to
play a fundamental role during the measurement process. As a matter of fact, a realistic measurement device sensible to a bounded
set $S$ of wave vectors $\bf k$ affects the states $|X\rangle$ and $|X_E \rangle$ in a dramatically different way after the
measurement since a measurement carried out on $|X\rangle$ leaves the modes ${\bf k} \notin S$ unaltered whereas the same
measurement on $|X\rangle_E$ profoundly changes the structure of the modes ${\bf k} \notin S$ causing, as usual, an irreversible
loss of information and notably of the propagation invariance.

\section{Generation and Detection of quantum X Waves}

The same difference between the two proposed quantum X waves also raises the problem of discerning which one (i.e. disentangled
or entangled) is obtained from a prescribed mechanism capable of generating a classical X wave. Since the disentangled X wave is
just the product of coherent states with different frequencies but with the same axicon angle, $|X\rangle$ is well expected to be
produced by a simple linear device, namely an axicon (as in \cite{Saari97}), exposed to a coherent non-monochromatic source like
a mode-locked laser. Conversely the entangled X wave is expected when considering nonlinear processes since it is well known that
frequency-conversion or parametric processes are accompanied by tight phase-matching condition able to provide the spectral
structure characterizing a classical X wave, as addressed by various authors with reference to various kind of nonlinearities (as
in \cite{Butkus05,Trillo02,Conti03bbis,Longhi04a}).

A viable scheme for detecting entanglement on a generated X-waves is offered by homodyne detection that select a specific mode,
eventually including a prism or a grating spatially separating the various angular frequencies. This correspond to consider the
projection of the state $|X_E \rangle$ onto a specific mode (at frequency $\omega_1$ and within a normalization constant)
\beq |X_E\rangle_{projected}= |\alpha_1\rangle +\Phi_E |0 \rangle \eeq
where $\Phi_E$ will in general be dependent on the specific X-wave or on its axicon angle.
Using homodyne detection (see \cite{YurkeALL}) and tuning the local field phase $\theta$ it is possible to unveil fringes
in the probability distribution of the detected current. As an example, for $\theta=-\arg(\alpha_1)+\pi/2$, the probability of
detecting the value $x$ of the output current $|\langle x|\alpha\rangle|^2$ is an oscillating function:
\beq |\langle x|\alpha\rangle|^2 = \frac{e^{-x^2}}{\sqrt{\pi}} \left[ 1+ |\Phi_E|^2+2|\Phi_E|\cos(\sqrt{2}|\alpha_1| x -\arg \Phi_E) \right]. \eeq
Note that the same reasoning with the state $|X\rangle$ yields $\Phi_E = 0$ so that oscillations in $|\langle x|\alpha\rangle|^2$ do
not appear if the original state is not entangled. More general results can be obtained by considering beam-splitters and generic
phases $\theta$, but their mathematical description is cumbersome and will be reported elsewhere.

In addition to the considered X waves, we discuss here the interesting possibility of turning a disentangled X wave into an
entangled one by letting the former (appropriately generated by an axicon) to pass through a nonlinear Kerr medium. In order to
avoid mathematical complications, we consider the typical spectrum of a broad band mode-locked laser containing $M$ harmonics
$\omega_m=m\omega_0$ ($m=1,2,...,M$) and we focus on the case where each mode is in a coherent state corresponding to a plane
wave propagating with the conical angle $\theta_0$. The disentangled X wave generated by the axicon is formed by the modes whose
wave vectors are such that $k_m=m\omega_0/c$ with $k_z=k_m/(1+\eta^2)$ and reads $|X\rangle=|\alpha_1 \rangle_1 |\alpha_2
\rangle_2 ...  |\alpha_M \rangle_M$ where $\alpha_m$ denotes the complex parameter of the coherent state with angular frequency
$\omega_m$. When travelling through a Kerr medium the various modes induce cross and self-phase modulation and the relevant and
well-known interaction Hamiltonian corresponds to that of a nonlinear oscillator (see \cite{YurkeALL}) and reads $\hat{H_I} =
\chi (\hbar/2) \sum_{m,p} \hat{n}_m \hat{n}_p$, where $\hat{n}_m= \hat{a}_m^\dagger \hat{a}_m$ is the photon number operator of
the mode $m$. If $t$ is the interaction time, the output state from the crystal $|X_{out}\rangle = e^{-\frac{i}{\hbar} \hat{H_I}
t} |X\rangle$ is given by
\beq |X_{out}\rangle  = e^{-\frac{1}{2} \sum_{r=0}^{M} |\alpha_r|^2} \sum_{n_1,n_2,...,n_M=0}^{\infty} e^{-\frac{i}{2}\chi t
\sum_{m,p=0}^{M} n_m n_p} \prod_{q=1,2,...,M} \frac{\alpha_q^{n_q}}{\sqrt{n_q !}}|n_q\rangle. \eeq
Note that, for $t=4 \pi / \chi$ the output state coincides with the incoming one whereas for $t=2 \pi / \chi$ the state
$|X_{out}\rangle$ is obtained from $|X\rangle$ after the replacement $\alpha_m \rightarrow-\alpha_m$ and the resulting
disentangled X wave state can be denoted as $|-X\rangle$. More interesting is the situation for $t=\pi/\chi$ since, defining
$\mathcal{N}=\sum_m n_m$ and exploiting the relation $e^{-i (\pi/2) \mathcal{N}^2}=[e^{-i\pi/4}+(-1)^\mathcal{N}e^{i\pi/4}]
/\sqrt{2}$, we obtain obtain
\begin{equation} \label{sc_state}
|X_{out}\rangle= \frac{1}{\sqrt{2}} \left( e^{-i\pi/4}|X\rangle + e^{i\pi/4}|-X\rangle \right).
\end{equation}
which is an entangled superposition of two disentangled quantum X waves travelling at the same velocity. The key point is here
that the interaction time $t$ coincides with the time spent by the classical X wave to pass through the nonlinear medium or $t =
D \cos \theta_0 / c$, $D$ being the nonlinear medium length. As for coherent states, the states periodically evolve with temporal
period $4 \pi/\chi$ and with spatial period $ 4 \pi c / (\cos \theta_0 \chi)$. As a consequence, by simply acting on the velocity
of the X-wave (e.g. by varying the axicon angle) it is possible to fine-tune the degree of entanglement of the two macroscopic
classical X-waves, and even to switching from classical states to Schr\"odinger-cat states. This is a not-trivial outcome of the
interplay between the classical interference process supporting X waves and the entanglement of coherent states. In addition, X
waves travel at a velocity which is different from the other linear waves so that, by using coincidences measurements, it is
straightforward to distinguish these states from a statistical mixture, and hence to point out their purely quantum properties as
described above.

\section{Conclusions}

In conclusion we have shown that classical electromagnetic X-waves may hide different degrees of quantum entanglement, depending
on their generation mechanism (exploiting a linear, a nonlinear medium or both of them). In this respect their velocity (i.e. the
axicon angle) plays a prominent role. Other interesting consequences of the X-waves structure arise when dealing with
interferometric setup, as for example that considered in \cite{Sanders92}. If a nonlinear medium is placed in one or both arms of
an interferometer, the output state is a ``progressive undistorted squeezed vacuum'' or entangled superposition of classical X
waves so that any detection scheme is affected by the axicon angle, and tuning among various quantum states can be attained, as
it will be detailed elsewhere. The quantum properties of X-waves can hence be exploited for free-space quantum communications,
highly sensible interferometers or quantum computing.

\begin{acknowledgments}
We acknowledge fruitful discussions with E. Del Re, S. Trillo, B. Crosignani and P. Di Porto.
\end{acknowledgments}

%\bibliography{IEEEabrv,ciattoni_quantumXwaves}

\end{document}